\documentclass[12pt,a4paper]{article}
\pdfoutput=1
\usepackage[a4paper,margin=1.5cm]{geometry}
\usepackage{setspace}
\usepackage{amsmath,amssymb}
\usepackage[T2A]{fontenc}
\usepackage[utf8x]{inputenc}
\usepackage[russian]{babel}
\usepackage{graphicx,xcolor}
\usepackage{gnuplottex}
\usepackage{bm}
\usepackage{caption}

\newcommand{\skp}{\vskip 1ex}
\newcommand{\mk}[1]{(M_{#1},\varkappa_{\parallel #1})}
\newcommand{\sgn}{\mathrm{sgn}}

\begin{document}
\captionsetup[figure]{labelfont={bf},labelformat={default},labelsep=period,name={Рис.}}

\begin{center}
{\bf О СВОЙСТВАХ РЕШЕНИЙ ДЛЯ МГД УДАРНЫХ ВОЛН В БЕССТОЛКНОВИТЕЛЬНОЙ ПЛАЗМЕ С ТЕПЛОВЫМИ ПОТОКАМИ}
\skp
{\bf Кузнецов В.Д., Осин А.И.\footnote{e-mail: osin@izmiran.ru}}

{\em ИЗМИРАН, Москва, Троицк }
\skp

{\bf ON THE PROPERTIES OF SOLUTIONS FOR MHD SHOCK WAVES IN COLLISIONLESS PLASMA WITH HEAT FLUXES}
\skp
{\bf Kuznetsov V.D., Osin A.I.}

{\em IZMIRAN, Moscow, Troitsk}
\skp
\end{center}

\section*{\centering\small Abstract}
\begin{center}
\begin{minipage}[c]{16cm}
The research is aimed at the study of the properties of solutions for parallel MHD shock waves in collisionless 
plasma with heat fluxes, using 8-moment MHD approximation. Restrictions have been established on the 
upstream shock parameters preserving physical meaning of the downstream plasma parameters. 

\vspace{1em}
{\bf Keywords:} collisionless plasma, heat fluxes, shock waves
\end{minipage}
\end{center}

\abstract{\noindent  
Обсуждаются особенности решений для параллельных МГД ударных волн в бесстолкновительной плазме с 
тепловыми потоками, полученных в рамках 8-моментного МГД приближения. 
Определены ограничения на область параметров плазмы 
перед фронтом ударной волны, в которой существуют решения с сохранением физического 
смысла параметров плазмы за фронтом ударной волны.
}

\section{Введение}
Бесстолкновительная плазма солнечного ветра, как показывают измерения на космических аппаратах 
\cite{Matteini2007,Stansby2018,Demars1990},  
характеризуется температурной анизотропией ($T_\parallel$ , $T_\perp$),  наличием тепловых потоков 
и турбулентностью. Ударные волны в космической плазме являются весьма распространенным  и 
часто наблюдаемым явлением. 
В анизотропной плазме ЧГЛ \cite{CGL} ударные волны изучались в ряде работ
 \cite{Lynn1967,AbrahamS1967,Neubauer1970,Hudson1977} , 
однако до сих пор не было попыток исследовать ударные волны в более сложных моделях 
бесстолкновительной плазмы, в частности, анизотропной плазмы с тепловыми потоками 
\cite{Oraevskii1968,Ramos2003}. 
В \cite{kvd-osin-pla-2018} было опубликовано 
решение соотношений на разрыве для параллельных ударных волн на основе 8-моментного приближения 
\cite{Oraevskii1968,Ramos2003}, определены области устойчивости плазмы перед фронтом 
ударной волны и за ним относительно ионно-звуковых возмущений.  
В \cite{kvd-osin-pla-2020} был рассмотрен вопрос о генерации ионно-звуковой,
шланговой и зеркальной неустойчивостей за фронтом параллельной ударной волны.  

В настоящей работе приводится анализ полученного решения, рассматриваются области  его 
существования и соответствующие ограничения на параметры ударных волн.

\section{Решение для параллельных ударных волн} 

В работе \cite{kvd-osin-pla-2018} приведены основные уравнения и соотношения Гюгонио на разрыве для 
параллельных ударных волн в анизотропной плазме с потоками тепла. 
Определим функцию сжатия $X$ плазмы в ударной волне
\begin{equation}
X \stackrel{def}{=} \frac{\rho_2}{\rho_1} 
\end{equation}
где индексом 1 обозначены параметры плазмы перед фронтом, 2 - за фронтом ударной волны. 
Из уравнения неразрывности следует, что 
\begin{equation}\label{ydef}
Y\stackrel{def}{=}\frac{u_2}{u_1}  = 1/X
\end{equation}
Полученное в \cite{kvd-osin-pla-2018} решение, описывающее параллельные ударные волны,  определяется 
функцией  $Y\mk{1}$, которая имеет вид:
\begin{equation}\label{Y}
Y_\pm\mk{1} = \frac{1}{2}\left(1+\frac{2}{M_1^2} \right) \pm \sqrt{D\mk{1}}
\end{equation}
где
\begin{equation}\label{D}
D\mk{1} =  \frac{1}{12} +\frac{1}{2M_1^4} - \frac{2\varkappa_{\parallel 1}} {3M_1^3}
\end{equation}
а давления и тепловые потоки за фронтом ударной волны определяются из следующих соотношений
\begin{align}
&\frac{p_{\parallel 2}}{p_{\parallel 1}} = P \label{ppara}\\
&\frac{p_{\perp 2}}{p_{\perp 1}}  = (1-\varkappa_{\perp 1}Z)  / Y  \label{pperp} \\
&\frac{S^\parallel_{\parallel 2}}{p_{\parallel 1}a_{\parallel 1}} = \varkappa_{\parallel 1}  + M_1(1-Y)(2-N)\\
&\frac{S^\perp_{\parallel 2}}{p_{\perp 1}a_{\parallel 1}} = \varkappa_{\perp 1} \frac{M_1^2-1}{N} \label{sperp} 
\end{align}
Здесь $M_1 = u_1/a_{\parallel 1}$ - безразмерная скорость потока плазмы перед фронтом ударной волны. 
Скорость потока $u_1$ отнесена к скорости тепловой волны 
$a_{\parallel 1} (a_{\parallel 1}^2=p_{\parallel_1}/\rho_1)$ из соображений удобства, так как число Маха $\mathcal M=u/c_{f,s}$ - 
скорость потока плазмы, отнесенная к скорости линейных ионно-звуковых волн, зависит от $\varkappa_{\parallel}$   -  
безразмерного потока  продольной тепловой энергии вдоль магнитного поля. 
Безразмерные тепловые потоки продольной ($S^\parallel_\parallel$) и поперечной ($S^\perp_\parallel$) 
тепловой энергии вдоль направления магнитного поля определяются следующим образом
\begin{equation}
\varkappa_{\parallel} = \frac{S^\parallel_{\parallel}}{p_{\parallel}a_{\parallel}} \quad , \quad 
\varkappa_{\perp} = \frac{S^\perp_{\parallel}}{p_{\perp}a_{\parallel}}
\end{equation}
Величины $P$, $N$ и $Z$ - функции параметров $\mk{1}$ : 
\begin{align}
&P\mk{1} = 1 + M_1^2 - M_1^2 Y \label{P} \\
&N\mk{1} = 2M_1^2Y - M_1^2 - 1 \label{N}\\
&Z\mk{1} = 2M_1(1-Y)/N \label{Z}
\end{align}

\begin{figure}[h]
\centering
\includegraphics[width=16cm]{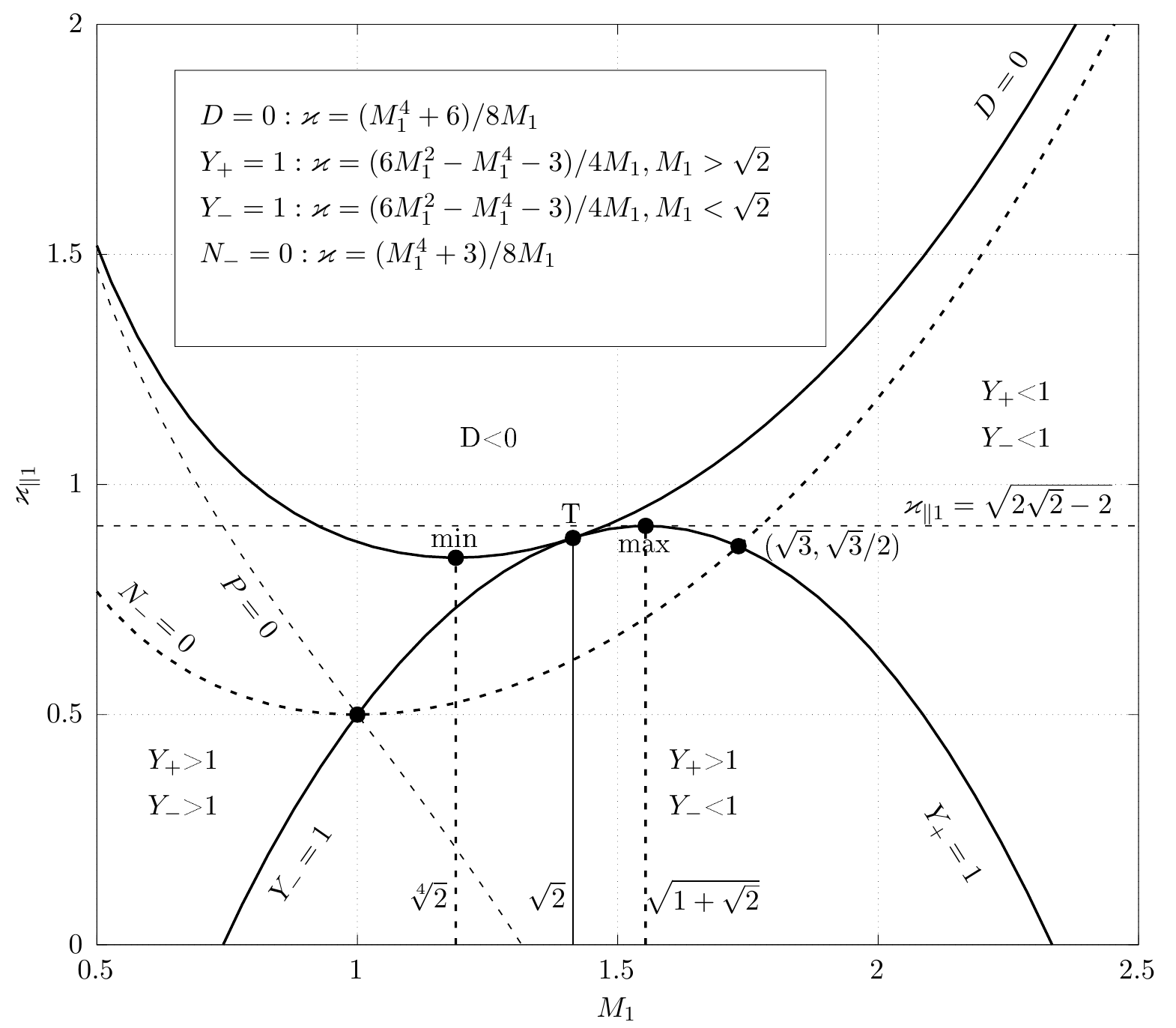}
\captionsetup{width=14cm}
\caption{Плоскость параметров $\mk{1}$ и основные кривые.} \label{fig:mk2}
\end{figure}

\section{Свойства решений для параллельных ударных волн}

Рассмотрим теперь более детально свойства различных функций, через которые выражаются 
параметры плазмы за фронтом ударной волны.

Различным знакам в (\ref{Y}) соответствуют два типа решений, причем, с учетом результатов о линейных 
быстрых и медленных ионно-звуковых волнах в анизотропной плазме с тепловыми потоками 
 \cite{Namikawa1981, Zakharov1988,Zakharov2015,Kuznetsov2009} можно заключить, что решение $Y_-$ описывает 
 медленные ударные волны, а $Y_+$  - как быстрые ($M_1>M_1^* = \sqrt{1+\sqrt{2}}$), так и медленные ($M_1<M_1^*$) 
 ударные волны (см. Рис.~(\ref{fig:mk2})). 

Так как $Y=\rho_1/\rho_2$, функция $Y\mk{1}$ должна быть вещественной, что ограничивает область определения решения на $\mk{1}$ областью
\begin{equation}
D\mk{1} \geqslant  0
\end{equation}
Вещественные решения существуют лишь для тех точек $\mk{1}$, которые лежат ниже кривой $K_{D0}$:
\begin{equation}
 \varkappa_{\parallel 1} = K_{D0}(M_1) = \frac{M_1^4+6}{8M_1}
\end{equation}
Кривая $K_{D0}$ имеет единственный минимум в точке $M_1=\sqrt[4]{2} \approx 1.19$, 
со значением в точке минимума $\varkappa_{\parallel 1} = 1/\sqrt[4]{2}  \approx 0.84$. 
Аналогичная ситуация, когда решение не существует из-за того, что некоторые величины за фронтом ударной волны становятся мнимыми, имеет место для ударной волны включения в изотропной  
МГД  (\cite{PolovinDem1987}, с. 123), а также ударной волны ионизации \cite{Taussig1965} .

Решение $Y_\pm\mk{1}$ представляет собой двузначную функцию, два листа значений которой расположены один над другим:
\begin{equation}
Y_+  \geqslant Y_-
\end{equation}
при этом решения совпадают ($Y_+=Y_-$) на кривой $D\mk{1} = 0$.
\begin{figure}[h]
\centering
\includegraphics[width=16cm]{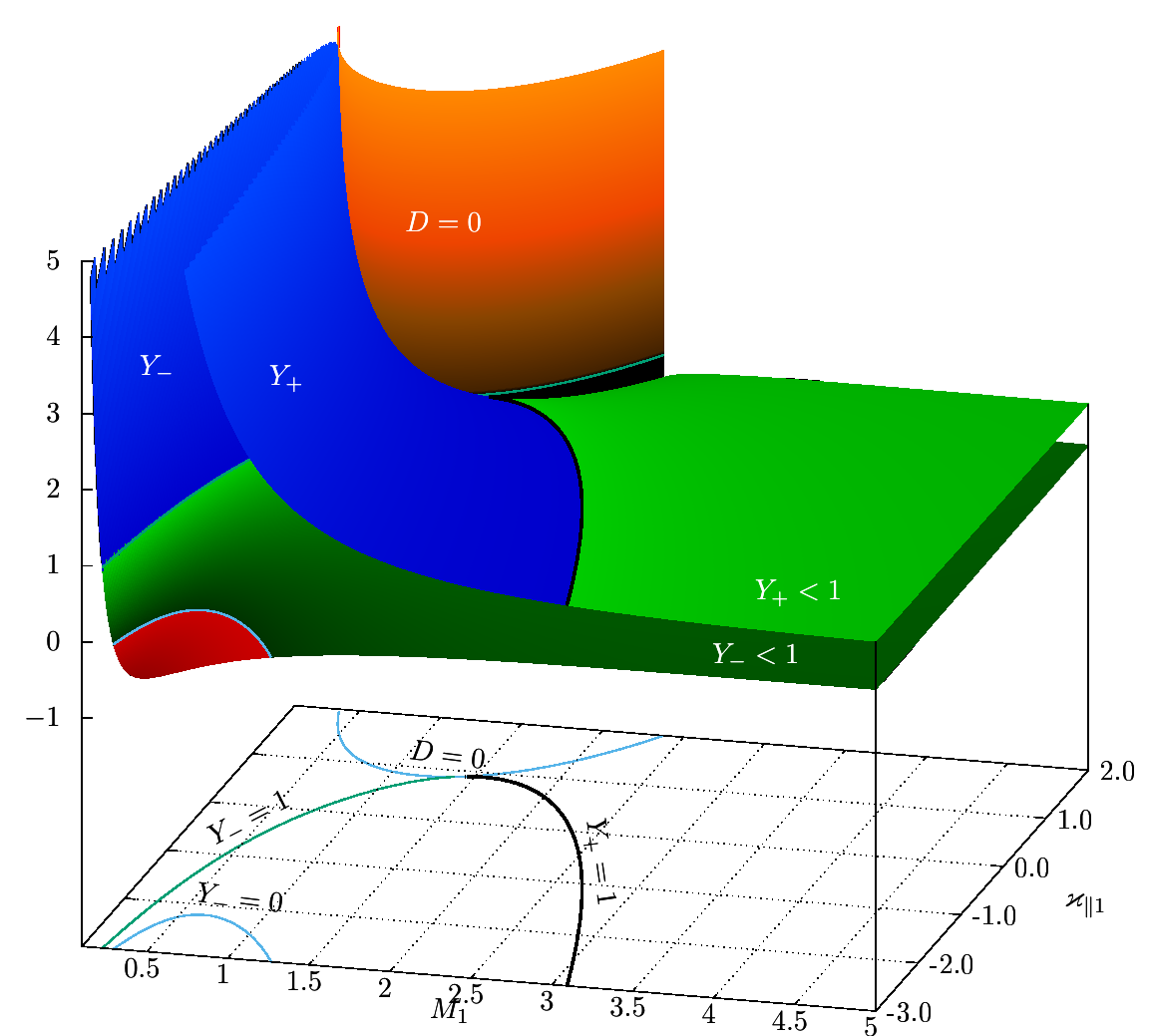}
\captionsetup{width=14cm}
\caption{Общий 3D вид решения $Y_\pm$ для параллельных ударных волн.}\label{mk2}
\end{figure}
Рассмотрим  дополнительные ограничения на возможные значения параметров $\mk{1}$ перед фронтом ударной волны 
обусловленные требованием сохранения физического смысла параметров плазмы за её фронтом. 

Прежде всего, функция $Y$, равная отношению плотностей, должна быть положительной. Для $Y_+$ это 
требование выполняется всегда,  в то время как $Y_-$ может принимать отрицательные значения для точек плоскости $\mk{1}$ , лежащих ниже кривой $K_{Y0}$
 \begin{equation}
 \varkappa_{\parallel 1} = K_{Y0}(M_1) = -\frac{M_1^4+6M_1^2+3}{4M_1}
\end{equation}
Кривая $K_{Y0}$ имеет единственный максимум в точке $\sqrt{\sqrt{2}-1}\approx 0.64$ и принимает в ней значение 
$-\sqrt{2\sqrt{2}+2} \approx -2.2$.

\begin{figure}[h]
\centering
\includegraphics[width=16cm]{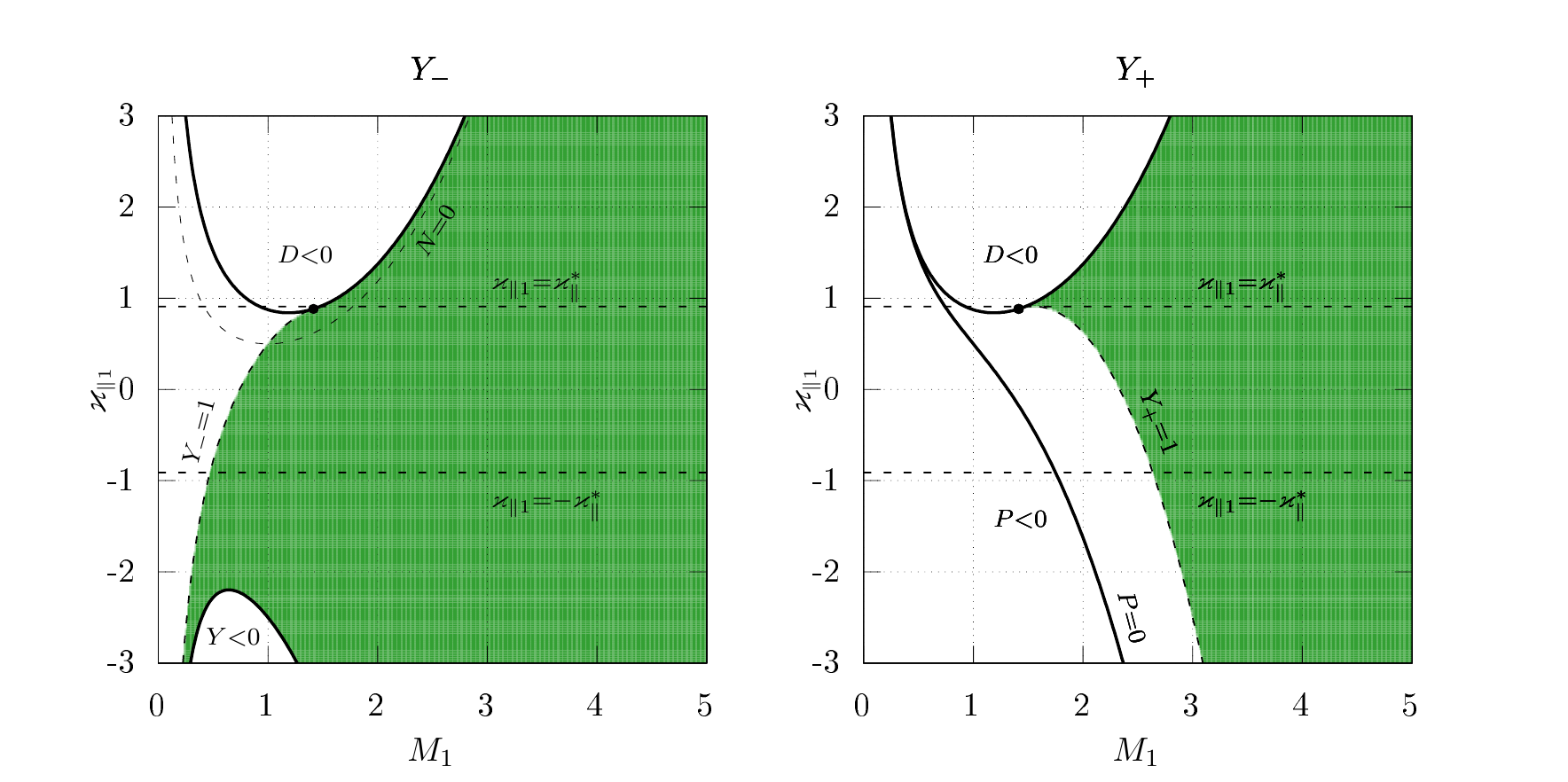}
\captionsetup{width=14cm}
\caption{Области определения решений $Y_-$ и $Y_+$. 
Закрашенным областям на рисунке соответствуют волны сжатия.
Области $D<0, Y<0, P<0$ не входят в область определения решений. 
}
\label{fig:ypm1}
\end{figure}

Кривая $K_{Y1}$, на которой $Y=1$, определяется с учетом (\ref{Y}):
\begin{equation}\label{KY1}
 \varkappa_{\parallel 1} = K_{Y1}(M_1) =  - \frac{M_1^4-6M_1^2+3}{4M_1}
\end{equation}
причем условиям $Y_{\pm}=1$ соответствуют различные ветви этой кривой на плоскости параметров $\mk{1}$:
\begin{align}
&Y_- = 1  : \quad \varkappa_{\parallel 1} = K_{Y1}(M_1)   \label{ym1}, 
\quad M_1 \leqslant  M_{T1} \\
&Y_+ = 1 : \quad \varkappa_{\parallel 1} = K_{Y1}(M_1)  \label{yp1},
\quad M_1 \geqslant M_{T1}
\end{align}
где $M_{T1} = \sqrt{2}$ -- точка касания кривых $K_{Y1}(M_1)$ и $K_{D0}(M_1)$. В соответствии с (\ref{ydef}) 
решениям $Y<1$ соответствуют ударные волны сжатия, $Y>1$ - волны разрежения (Рис~\ref{fig:ypm1}). Точки на 
кривой $Y_\pm=1$ соответствуют ударным волнам бесконечно малой интенсивности, для которых 
значения всех параметров плазмы за фронтом ударной волны равны их значениям перед фронтом. 

Продольное давление $p_\parallel$ также должно быть положительным.  Из (\ref{Y}),(\ref{D}),(\ref{P}) получим
\begin{equation}\label{PD}
P\mk{1} = M_1^2 \left( \frac{1}{2} \mp \sqrt{D} \right)
\end{equation}
Для $Y_-$ функция $P$ всегда положительна. Для  $Y_+$ , как следует из (\ref{PD}), $P>0$  если  $D<1/4$, то есть   
для точек $\mk{1}$ лежащих правее кривой $K_{P0}$:
\begin{equation}
\varkappa_{\parallel 1} = K_{P0}(M_1) =  \frac{3-M_1^4}{4M_1}
\end{equation}
Эта граница проходит в области параметров $\mk{1}$, для которых решению $Y_+$ соответствуют 
ударные волны разрежения (Рис.~\ref{fig:ypm1}).

Дополнительные ограничения на решения следуют из требования положительности поперечного 
давления $p_\perp$ за фронтом ударной волны, а также конечности поперечного давления  и 
потока поперечной тепловой энергии $S^\perp_\parallel$.
Здесь будет необходимо рассмотреть свойства функции (\ref{N}). 
На кривой $K_{N0}$, определяемой условием $N\mk{1}=0$, поперечное давление и поперечный 
тепловой поток за фронтом ударной волны  
при наличии отличного от нуля поперечного теплового потока перед фронтом ($\varkappa_{\perp 1} \ne 0$), 
как это следует из (\ref{pperp}),(\ref{sperp}),(\ref{Z}),  обращаются в бесконечность.
Анализ показывает, что функция $N\mk{1}$ может обращаться в ноль только для медленных ударных волн ($Y_-$) : 
\begin{equation}
N=0 \Leftrightarrow 1  - 2M_1^2\sqrt{D} = 0
\end{equation}
Откуда (только для $Y_-$) следует: 
\begin{equation}
 \varkappa_{\parallel 1} = K_{N0}(M_1) = \frac{M_1^4+3}{8M_1}
\end{equation}
Это - кривая, лежащая всюду ниже $D=0$ ($K_{N0}<K_{D0}$), т.к.
\begin{equation}
 \frac{M_1^4+3}{8M_1}  <  \frac{M_1^4+6}{8M_1} 
\end{equation}
Минимум $K_{N0}$ находится в точке $M_1=1$  (Рис. \ref{fig:ypm1})
\begin{equation}
\min_{M_1} K_{N0} = K_{N0}(1) = \frac{1}{2}
\end{equation}
Вернемся к анализу знака $p_{\perp 2}$. Из (\ref{pperp}), с учетом положительности  $Y,M_1$,  следует:
\begin{equation}
p_{\perp 2} < 0 \Leftrightarrow \varkappa_{\perp 1}Z > 1
\end{equation}
что имеет место, если знаки $\varkappa_{\perp 1}$ и $Z$ совпадают и поток поперечной тепловой энергии
$\varkappa_{\perp 1}$ достаточно велик. При этом
\begin{equation}
\sgn(Z) = \sgn\left(\frac{1-Y}{N}\right)
\end{equation}
Функции $K_{N0}, K^-_{Y1} (Y_-= 1)$ пересекаются в точке ($1,1/2$). Для знака потока поперечной тепловой энергии за фронтом 
ударной волны из (\ref{sperp}) имеем
\begin{equation}\label{sgnsperp}
\sgn(S^\perp_{\parallel 2})  = \sgn( S^\perp_{\parallel 1}) \; \sgn\left(\frac{M_1-1}{N} \right)
\end{equation}


\section{Свойства отображения $\mk{1} \rightarrow \mk{2}  $}

Функции $Y_\pm$ задают отображения $F^\pm_Y: \mk{1}\to\mk{2}$ точек пространства параметров $\mk{1}$ (перед фронтом ударной волны) в пространство параметров $\mk{2}$ (за фронтом ударной волны). Отображение задается парой функций
$F_M\mk{1}$ , $F_\varkappa \mk{1}$:
\begin{eqnarray}
M_2 = F_M\mk{1} = M_1\sqrt{ \frac{Y}{1+M_1^2-M_1^2Y} } &=& M_1 \sqrt{\frac{Y}{P}}  \label{m2} \\
\varkappa_{\parallel 2} = F_\varkappa\mk{1}= \frac{\varkappa_{\parallel 1}+M_1(1-Y)(2 - N)}  
{\sqrt{Y(1+M_1^2-M_1^2Y)^3}}  &=& \frac{\varkappa_{\parallel 1}+M_1(1-Y)(2 - N)}  {\sqrt{YP^3}}  \label{k2}
\end{eqnarray}
где $Y_\pm$ определяется выражениями (\ref{Y}),(\ref{D}). Областью определения отображения $F_Y^-$, поскольку в этом случае $P>0$, является множество точек пространства параметров $\mk{1}$ 
\begin{equation}\label{defm}
{\mathcal D(Y_-)} : D\mk{1} \geqslant 0 \cap Y_-\mk{1}>0
\end{equation}
Областью определения отображения $F_Y^+$ (здесь $Y_+>0$) являются точки плоскости $\mk{1}$, определяемые выражением 
\begin{equation}\label{defp}
{\mathcal D(Y_+)} : D\mk{1} \geqslant 0  \cap P\mk{1}>0
\end{equation}
\begin{figure}[h]
\centering
\includegraphics[width=16cm]{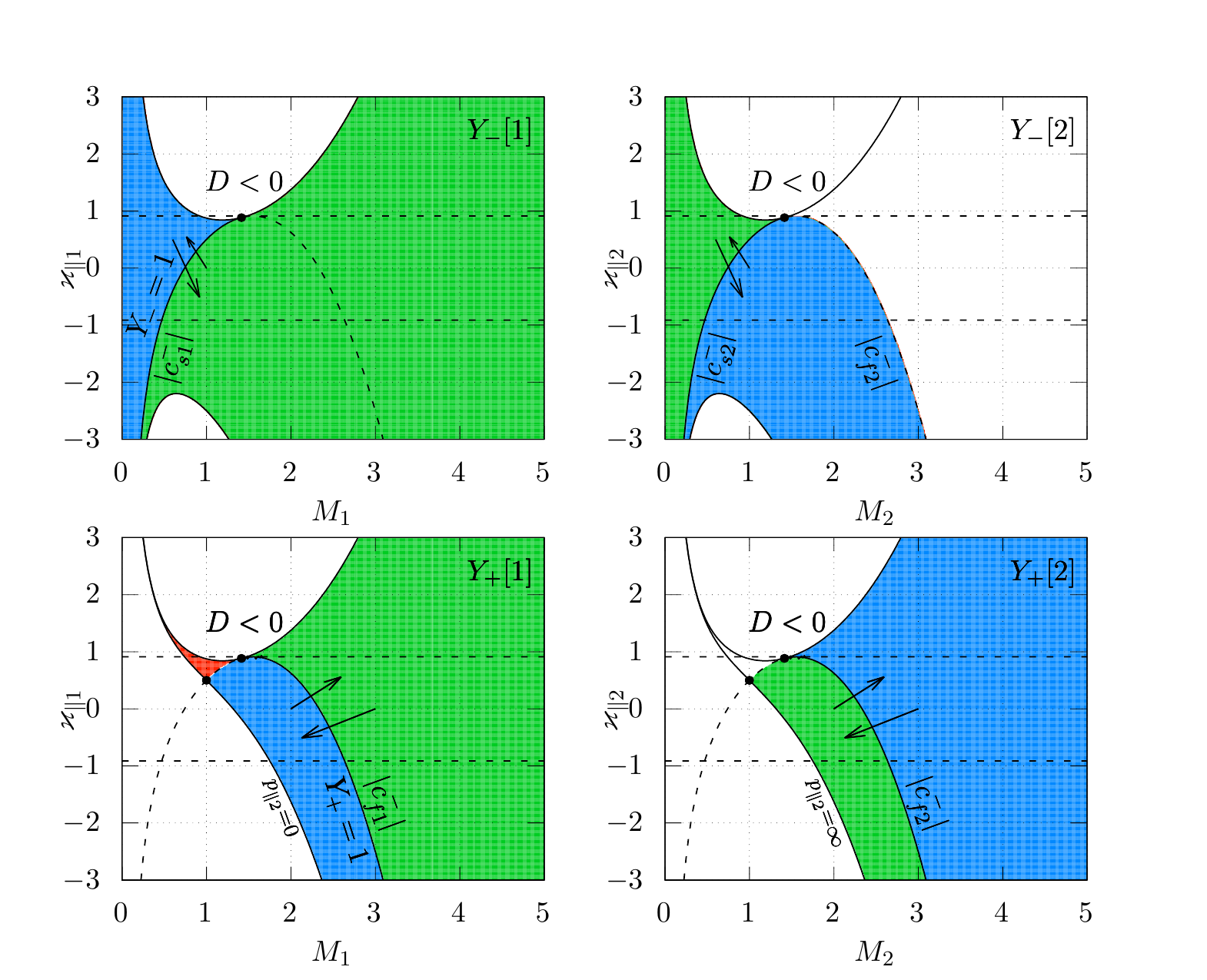}
\captionsetup{width=14cm}
\caption{Области определения (рисунки слева) и значений (рисунки справа) отображения $F^\pm_Y$ пространства 
параметров $\mk{1}$ в пространство параметров $\mk{2}$.} \label{fig:map}  
\end{figure}
Области определения отображений $F^\pm_Y$ приведены на Рис.~\ref{fig:map} (закрашены, рисунки слева). 

Областью значений этих отображений являются определенные подобласти пространства параметров $\mk{2}$ 
(рисунки справа).  При этом кривая $K_{D0}(D=0)$ (в этом случае $Y_+=Y_-$) отображается на 
$K_{Y1}$, левая ветвь ($M_1<M_{1T}$) - на правую ветвь $K_{Y1}$ ($K_{Y1}^+$), 
правая ($M_1>M_{1T}$) - на часть левой ветви $K_{Y1} (K_{Y1}^-)$, где $M_1>1$.

На рисунках также обозначены кривые, соответствующие взятым по модулю безразмерным
($c^-_{f,s} = a^-_{f,s}/a_\parallel$) скоростям быстрых ($c^-_f$) и медленных ($c^-_s$) линейных 
ионно-звуковых возмущений, распространяющихся в отрицательном направлении (в направлении 
невозмущенного потока плазмы перед фронтом ударной волны). 

Не все параметры $\mk{2}$ имеют соответствующий прообраз на плоскости $\mk{1}$. Эти параметры 
за фронтом ударной волны не реализуются ни при каких значениях параметров $\mk{1}$ перед фронтом. 
Также, не все точки на $\mk{2}$ имеют единственный прообраз, т.е. отображение $F_Y$ не является 
инъективным.  Это означает, что некоторым состояниям за фронтом ударной волны $\mk{2}$ 
соответствуют два  решения с различными $\mk{1}$.

Для тех точек $\mk{2}$, для которых существует обратное отображение, его легко получить 
из соотношений на разрыве (соотношений Гюгонио). Поскольку эти соотношения симметричны 
относительно состояний перед (1) и за (2) фронтом ударной волны, разрешение этих уравнений  
относительно состояния за фронтом приводит к идентичным функциональным выражениям 
с заменой индексов (1)  и (2) и $X$ вместо $Y$.

В результате получаем обратное отображение $G_X\mk{2}=F^{-1}_{X}\mk{2}$, такое, что
$G_X(F_Y\mk{1}) = id\mk{1}=\mk{1}$, а $\varkappa_{\parallel 1}$ и $M_1$ выражаются следующим образом
\begin{equation}\label{m1}
M_1 = M_2\sqrt{ \frac{X}{1+M_2^2-M_2^2X} }
\end{equation}
\begin{equation}\label{k1}
\varkappa_{\parallel 1} =\frac{\varkappa_{\parallel 2}+M_2(1-X)(2-N)}
{\sqrt{X(1+M_2^2-M_2^2X)^3}}
\end{equation}
где 
\begin{equation}\label{x2roots}
X_{\pm} = \frac{1}{2}\left(1+\frac{2}{M_2^2}\right) \pm 
\sqrt{ \frac{1}{12}+\frac{1}{2M_2^4}-\frac{2\varkappa_{\parallel 2}}{3M_2^3} }
\end{equation}
Для определения области значений $F_Y$ определим на $\mk{2}$ множество точек, для которых $F_Y(G_X) = id$,
то есть тех точек $\mk{2}$, которые при последовательном применении отображений $G_X: \mk{2} \to \mk{1} $ и
 $F_Y: \mk{1}\to\mk{2}$ возвращаются в исходную точку. Результат графического решения этой задачи 
 представлен на Рис.~\ref{fig:map}.
 
Для дальнейшего анализа необходимо пояснить смысл линий на $\mk{1}$ и $\mk{2}$ нанесенных пунктиром.
Из дисперсионного уравнения для малых одномерных возмущений 
следует уравнение для безразмерных скоростей  ($y=\omega/ka_\parallel$) малых ионно-звуковых возмущений вдоль магнитного поля \cite{Namikawa1981,kvd-osin-pla-2018}
\begin{equation}
y^4-6y^2-6\varkappa_\parallel y + 3 = 0
\end{equation}
откуда
\begin{equation}\label{kappay}
\varkappa_\parallel = -\frac{y^4-6y^2+3}{4y}
\end{equation}
Можно заметить, что это уравнение совпадает с (\ref{KY1}), где используется переменная 
$M=u/a_{\parallel}$ -- безразмерная скорость потока перед фронтом ударной волны (скорость ударной волны), 
тогда как в (\ref{kappay}) -- безразмерная фазовая скорость волн и 
обе величины отнесены к скорости тепловой волны $a_\parallel$.  Это совпадение отражает известный факт о том, 
что ударные волны бесконечно малой интенсивности распространяются со скоростью соответствующих 
малых (быстрых или медленных ионно-звуковых) возмущений. 

При описании и анализе различных решений важно иметь возможность сравнивать скорости ударных 
волн со скоростями малых возмущений ($\mathcal M = u/a$), прежде всего со скоростями малых 
возмущений, распространяющихся против потока плазмы в системе отсчета, связанной с ударной волной. 
При этом нельзя забывать, что в рассматриваемой модели плазмы скорости 
линейных волн, вообще говоря (при $\varkappa_{\parallel} \ne 0$), несимметричны относительно 
направления  теплового потока.
 
Уравнение (\ref{kappay}) имеет 4 различных вещественных решения $y_{1-4}$ при $|\varkappa_{\parallel}|<\varkappa^*$.
Обозначим эти решение как $c_{f,s}^\pm$ - медленные (s) и быстрые (f) линейные волны, распространяющиеся  в положительном (+) и отрицательном (--) направлении.  
Уравнение (\ref{kappay}) позволяет отобразить кривые линейных скоростей $c_{f,s}^\pm$ на ($M,\varkappa_{\parallel}$), 
причем эту схему можно рассматривать как в плазме перед фронтом  ударной
волны ($M_1,\varkappa_{\parallel 1}$), так и за фронтом ($M_2,\varkappa_{\parallel 2}$).    
\begin{figure}[h]
\centering
\includegraphics[width=14cm]{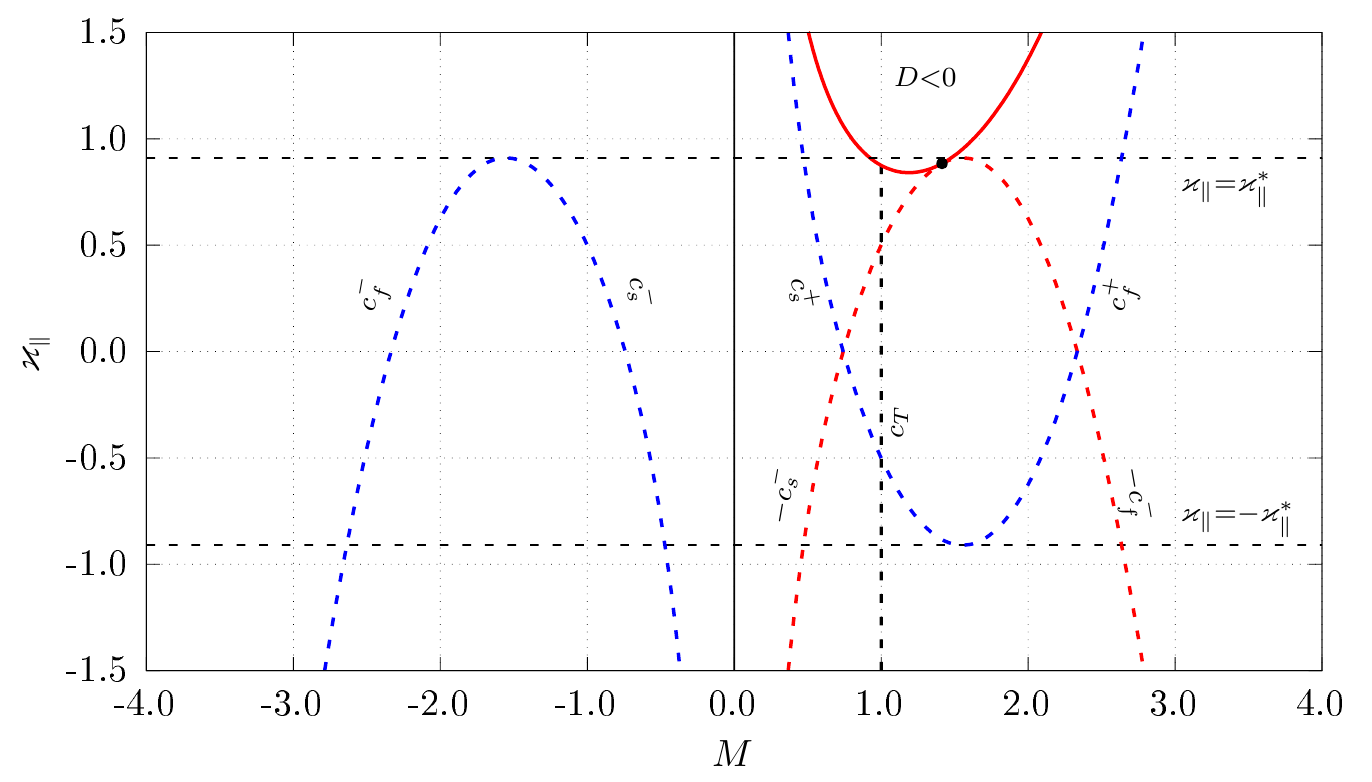}
\captionsetup{width=14cm}
\caption{Фазовые скорости линейных  ионно-звуковых волн $c_{f,s}^\pm$. Скорость тепловой волны $c_T=1 \;(a_T=a_\parallel)$. Красная пунктирная линия : $-c^-_{f,s}$ .}
\label{fig:cfs}   
\end{figure}

Если отобразить на $\mk{}$ взятые по абсолютной величине безразмерные скорости идущих против потока плазмы 
линейных волн, то $|c_{f,s}^-|$  совпадет с $K_{Y1}$ (Рис.~\ref{fig:map}, отмечены пунктиром, красным цветом). 
Для точек, лежащих ниже (``внутри'') указанной кривой, образованной ветвями $\varkappa_\parallel =|c_{f,s}^-|$,
справедливы неравенства
\begin{equation}
  \mathcal M_s^- > 1\quad , \quad \mathcal M_f^- < 1
\end{equation}
здесь $M_{f,s}^-$ - ``быстрое'' и ``медленное'' числа Маха, $\mathcal M_{f,s}^- = M/|c_{f,s}^-|$. Соответственно,
для точек, лежащих левее кривой $|c_s^-|$ 
\begin{equation}
  \mathcal M_s^- < 1\quad , \quad \mathcal M_f^- < 1
\end{equation}
а для точек, лежащих правее кривой $|c_f^-|$ 
\begin{equation}
  \mathcal M_s^- > 1\quad , \quad \mathcal M_f^- > 1
\end{equation}

Горизонтальными пунктирными линиями на Рис.~\ref{fig:ypm1},\ref{fig:map},\ref{fig:cfs} отмечены границы области ионно-звуковой 
неустойчивости $|\varkappa_\parallel| = \varkappa^* = \sqrt{2\sqrt{2}-2}\approx 0.91$ \cite{Namikawa1981,kvd-osin-pla-2018}.

Для ударных волн $Y_-$ вся область определения $\mathcal D(Y_-)$(\ref{defm}) 
отображается на $\mk{2}$ без точек, лежащих выше $K_{D0}$, ниже $K_{Y0}$ и правее $|c_{f2}^-|$. 
При этом точки $\mk{1}$, лежащие ниже $K_{D0}$ и правее $K^-_{Y1}$,  
соответствуют (медленным) ударным волнам сжатия. За фронтом, на $\mk{2}$ им соответствуют точки,
лежащие ниже $K_{D0}$ и левее $|c_{s2}^-|$, то есть поток за фронтом медленной ударной волны 
сжатия $Y_-$ дозвуковой ($\mathcal M^-_{s 2}<1, \mathcal M^-_{f 2}<1$).
Точки, лежащие ниже $K_{D0}$ и левее $K^-_{Y1}$ соответствуют медленным ударным волнам разрежения.
За фронтом им соответствуют точки, лежащие ниже  кривых $|c_{s 2}^-|$, $|c_{f 2}^-|$, то есть течение за
фронтом медленной ударной волны разрежения сверхзвуковое ($\mathcal M_{s 2}^->1$) по отношению 
к медленной и дозвуковое ($\mathcal M_{f 2}^-<1$) по отношению к быстрой ионно-звуковой волне за фронтом.

\begin{figure}[h]
\centering
\includegraphics[width=14cm]{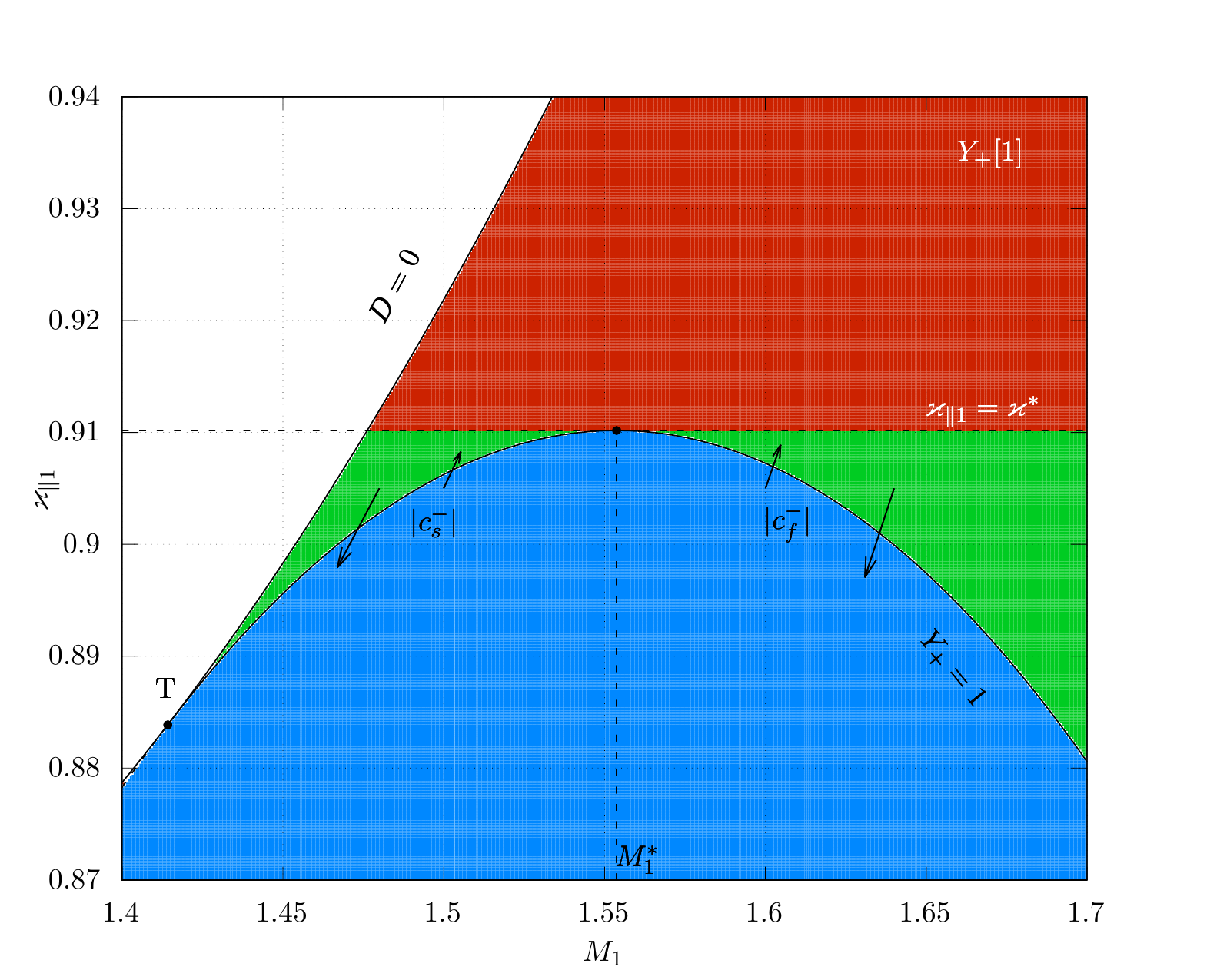}
\captionsetup{width=14cm}
\caption{Плоскость $(M_1,\varkappa_{\parallel 1})$ в окрестности точки касания $T$ кривых $K^+_{Y1}\;(Y_+=1)$ и $K_{D0}\;(D=0)$. Фазовые скорости линейных  ионно-звуковых волн $c_{f,s}^-$  совпадают с $K^+_{Y1}$. Стрелками обозначены переходы через фронт ударной волны $\mk{1} \to \mk{2}$. Зеленым цветом обозначены волны сжатия, синим - волны разрежения, красным - область ионно-звуковой неустойчивости.}
\label{fig:mapt}   
\end{figure}

Для ударных волн $Y_+$ область определения решений $\mathcal D(Y_+)$ отображается на аналогичную 
область  пространства параметров $\mk{2}$, кроме точек, лежащих ниже $K_{D0}$, левее $|c^-_{s 2}|$ и 
правее $K_{P0}$ (узкая область параметров вдоль $D\mk{2}=0$, Рис.~\ref{fig:map} ). Точки $\mk{1}$, лежащие 
ниже $K_{D0}$ и правее $|c^-_{f 1}|$, соответствуют быстрым ударным волнам сжатия, лежащие 
левее $|c^-_{s1}|$ -- медленным ударным волнам сжатия $Y_+$. 
За фронтом, на $\mk{2}$ (переход обозначен на Рис.~\ref{fig:mapt} стрелками), им соответствуют точки, 
лежащие ниже $|c_{s 2}^-|,|c_{f 2}^-|$ и правее 
$K_{P0}(M_2)$. Поток плазмы перед фронтом быстрой ударной волны сжатия сверхзвуковой, 
$\mathcal M^-_{f1}>1$, за фронтом -- дозвуковой, $\mathcal M_{f 2}^-<1$\; (но $\mathcal M_{s 2}^->1$).
Медленная ударная волна сжатия $Y_+$ является особенной в том смысле, что в этом случае поток  
плазмы перед фронтом дозвуковой ($\mathcal M^-_{s1}<1$), а за фронтом  -- 
сверхзвуковой ($\mathcal M_{s 2}^->1$), Рис.~(\ref{fig:mapt}), хотя скорость потока плазмы при этом падает. 
Такая особенность обусловлена наличием зависимости скороcтей линейных волн от величины теплового потока $\varkappa_\parallel$. 

Область быстрых ударных волн разрежения ($D\geqslant 0 \cap P>0 \cap Y_+>1$) распадается 
на две подобласти с $\mathcal M_{s}^->1$ и $\mathcal M_{s}^-<1$ (Рис.~\ref{fig:map}), причем обе указанные подобласти 
отображаются на область $D\geqslant 0 \cap M_2>|c_{f 2}^-|$. Таким образом, определенное состояние  $\mk{2}$  за 
фронтом быстрой ударной волны разрежения реализуется в двух различных решениях c различными 
состояниями перед фронтом ударной волны $\mk{1}$.   

Здесь следует сделать пояснение, что не все описанные решения имеют физический смысл (физически 
реализуемы). 
Во-первых, некорректно рассматривать решения для плазмы, состояние которой удовлетворяет условиям 
развития ионно-звуковой неустойчивости $|\varkappa_{\parallel 1}|>\varkappa^*$ 
(Рис~1. в \cite{kvd-osin-pla-2018}, \cite{kvd-osin-pla-2020}). 
Во-вторых, для продолжительного существования стационарной структуры ударных волн необходимо, чтобы 
и за фронтом состояние было устойчивым, $|\varkappa_{\parallel 2}|>\varkappa^*$. Вопросу возникновения ионно-звуковой, зеркальной и шланговой неустойчивости плазмы за фронтом  ударных волн посвящена работа \cite{kvd-osin-pla-2020}.
Кроме того, необходимо исследовать вопрос эволюционности указанных решений. Предварительные 
результаты такого исследования (готовятся к печати) указывают на то, что, в частности,  ударные волны разрежения 
неэволюционны.

\section{Заключение}
В работе осуществлен анализ полученных ранее решений для параллельных ударных волн  в бесстолкновительной 
плазме на основе 8-моментного МГД приближения.  
Изучена структура функции $Y\mk{1}$,  определяющей решения и задающей отображение пространства параметров 
перед фронтом ударной волны $\mk{1}$ в пространство параметров за фронтом $\mk{2}$.
Проанализированы свойства данного отображения, его область определения и область значений.  
Установлены ограничения на определяющие решения параметры $\mk{1}$,  при которых существуют решения  в виде стационарных ударных волн. 
Полученные ограничения обеспечивают вещественность всех физических величин, а также положительность плотности, скорости и давления за фронтом ударной волны.  Проведенный детальный анализ свойств различных функций, реализующих описанные решения,  предоставляет возможность для их дальнейшего изучения, в частности, 
рассмотрения вопроса об эволюционности параллельных ударных волн в плазме с тепловыми потоками.


\bibliographystyle{ugost2008l}
\bibliography{bibshort,mybib}

\end{document}